\documentclass[aps,pre,english,showpacs,showkeys,preprintnumbers,twocolumn,floatfix,nofootinbib,10pt]{revtex4-2}
\usepackage{eurosym}
\usepackage{amssymb}
\usepackage{graphicx,color}
\usepackage{epsfig}
\usepackage{rotating}
\usepackage[T1]{fontenc}
\usepackage{latexsym,epsfig}
\usepackage[latin9]{inputenc}
\usepackage{amsfonts}
\usepackage{dcolumn}
\usepackage{bm}
\usepackage{babel}
\usepackage{hyperref}
\usepackage{amsmath,mathrsfs}

\begin{document}

\title{Affinities and disagreements between Mean-field and short-range
critical dynamics}
\author{Roberto da Silva}
\address{Instituto de F{\'i}sica, Universidade Federal do Rio Grande do Sul,
UFRGS}

\begin{abstract}
In this work, we explore some interesting details of the time-dependent
regime of the long-range systems under mean-field approximation in
comparison with the critical dynamics of the short-range systems. First, we
discuss some mechanisms of the initial anomalous behavior of the
magnetization via two-dimensional Monte Carlo simulations to after compare
with results from Mean-field simulations in both: spin 1/2 (Ising) and spin
1 (Blume Capel model) Ising models. The distinction between critical and
tricritical points is also investigated. For a complete analysis, we
performed short-time simulations in the mean-field regime to determine the
critical temperatures optimizing power laws and the critical exponents of
the different points, which were independently calculated, i.e. without
using previous critical exponents estimates from literature/theory. Our
investigations corroborate analytical results here also developed.
\end{abstract}

\maketitle



\section{Introduction}

\label{sec:introduction}

The studies in statistical mechanics can be essentially divided into two
parts: equilibrium (or stationary regime for non-Hamiltonian systems) and
nonequilibrium one \cite{Salinas-Reichl}. A huge number of studies were
dedicated to the equilibrium branch, and in this context, the long-range
(LR) mean-field (MF) regime approximation for such systems was/is explored
in several models, however, we believe that several points in this regime
deserve better attention when explored in nonequilibrium regime since some
points relatively well established in short-range (SR) systems as the
relaxation of the magnetization in MF, did not have suitable exploration in
the literature to the best of our knowledge.

Considering for example a description with time-dependent (TD) Monte Carlo
(MC) simulations (TDMC) in the MF regime, it could be interesting to
investigate the dynamic behavior of systems in MF approximation as well as,
investigating the possible similarities/differences in this study in
comparison with regular TDMC simulations of spin systems with SR
interactions, for instance, in two or three-dimensional lattices.

Ising-like Hamiltonians can be simply generalized as written in the
following:

\begin{equation}
\mathcal{H}=-J\sum\limits_{\left\langle i,j\right\rangle }\sigma _{i}\sigma
_{j}+D\sum\limits_{i=1}^{N}\sigma _{i}^{2}-H\sum\limits_{i=1}^{N}\sigma _{j}
\label{Eq:short_range_hamiltonian}
\end{equation}%
where if $D=0$ and $\sigma _{j}=\pm 1$ (spin 1/2) one has the standard Ising
model, while for $D\geq 0$ (anisotropy term) and $\sigma _{j}=0,\pm 1$ (spin
1) one has the known Blume-Capel model. Here $H$ is the external field that
couples with each spin and $\left\langle i,j\right\rangle $ denotes that sum
is taken only over the nearest neighbors in a $d$-dimensional lattice.

A mean-field approximation considers that each $i$-th spin $\sigma _{i}$
interacts with a magnetic \textquotedblleft cloud\textquotedblright\
represented by the average magnetization $\xi _{i}=$ $\frac{1}{N}%
\sum\limits_{j=1,j\neq i}^{N}\sigma _{j}$. Each spin, in the supposed
original lattice where it is inserted, is linked to other $z=2^{d}$
neighbors, the number of links in the lattice is $\frac{Nz}{2}$ since one
has to count $\frac{z}{2}$ links for each read spin by avoiding repeated
links. Thus in this approximation, the interacting term $\mathcal{H}%
_{int}=-J\sum\limits_{\left\langle i,j\right\rangle }\sigma _{i}\sigma _{j}$
must be replaced \ in the mean-field approximation by%
\begin{equation*}
\begin{array}{lll}
\mathcal{H}_{int}^{(MF)} & = & -\frac{Jz}{2}\sum\limits_{i=1}^{N}\sigma
_{i}\xi _{i} \\ 
&  &  \\ 
& = & -\frac{Jz}{2N}\sum\limits_{i=1}^{N}\sigma _{i}\sum\limits_{j=1,j\neq
i}^{N}\sigma _{j} \\ 
&  &  \\ 
& \approx & -\frac{Jz}{2N}\sum\limits_{i=1}^{N}\sum\limits_{j=1}^{N}\sigma
_{i}\sigma _{j}%
\end{array}%
\end{equation*}

Finally, one has that mean-field Hamiltonian is given by 
\begin{equation}
\mathcal{H}^{(MF)}=-\frac{Jz}{2N}M^{2}-HM+D\sum\limits_{i=1}^{N}\sigma
_{i}^{2}  \label{Eq:long_hange_hamiltonian}
\end{equation}%
where $M=\sum\limits_{i=1}^{N}\sigma _{i}$ is the magnetization of the
system. For our aims $H=0$ from here.

At this point, we raise an important question in this manuscript: What are
the similarities and differences between the relaxation of SR spin systems
(hamiltonian from Eq. \ref{Eq:short_range_hamiltonian}) and LR-MF spin
systems (that one from Eq \ref{Eq:long_hange_hamiltonian})? For that, we
must remember some points about the relaxation of spin systems with SR
interactions. Such systems, initially at high temperature, and therefore
highly disordered, (with very small initial magnetization $m_{0}<<1$), when
suddenly placed in contact with a thermal reservoir at critical temperature $%
T_{c}$, tend to initially mimic the MF characteristic correlations, and the
correlations must present a kind of \textquotedblleft
inertia\textquotedblright\ until reaching the regime of SR correlations.

Such tendency leads to an initial anomalous behavior which is not exactly
the same for critical (see for example \cite{Zheng1998,SilvaPLA2002}) and
tricritical points \cite{SilvaPRE2002,SilvaTRICPC2013}. \ At this same
initial condition, for critical points, the theory of Jansen, Schaub, and
Schmittmann \cite{Janssen1989} and the Monte Carlo (MC) simulations from
Zheng \cite{Zheng1998} predict a crossover between two power laws: the first
one is exactly an anomalous increase of the magnetization from the initially
disordered state characterized for an exponent $\theta >0$, followed by a
decrease of the magnetization which occurs when the system reaches a
reasonable ordering state and in this case with an exponent $\lambda <0$.
Finally, the system remains decaying but in this case exponentially after
reaching the thermodynamic equilibrium.

However tricritical points present peculiar aspects when initially prepared
with a small magnetization as predicted by Oerding and Jansen \cite%
{Janssen1994}. The anomalous initial behavior in two dimensions is
characterized by a power-law with $\theta <0$. Here, only for illustration,
we performed standard TDMC simulations according to Metropolis dynamics, the
magnetization $m(t)=\frac{1}{N}\left\langle \sum_{i=1}^{N}\sigma
_{i}\right\rangle $ averaged over different $N_{run}=15000$ runs, for the
two-dimensional Blume-Capel model, for one critical point (anisotropy $D/J=0$
and $k_{B}T_{C}/J=1.6950$) and the tricritical one ($D/J=1.9655$ $%
k_{B}T_{C}/J=0.610$), for $L=160$ for several values of initial
magnetization (fixed but randomly established in the beginning of each run).
Please, for details of how to perform these simulations, see for example 
\cite{Zheng1998,SilvaPRE2002}.

\begin{figure}[h]
\begin{center}
\includegraphics[width=1.0\columnwidth]{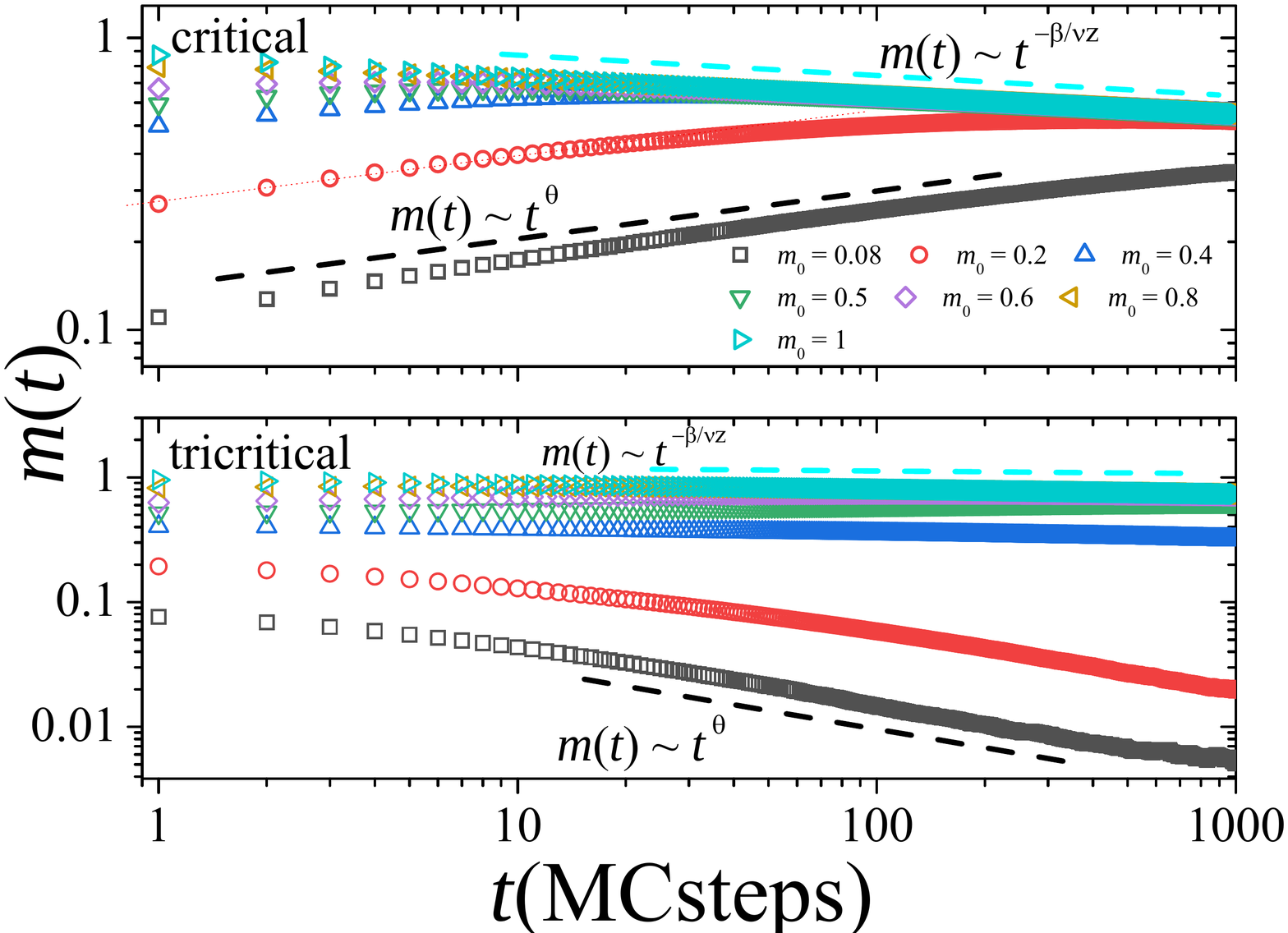}
\end{center}
\caption{Time evolutions of magnetization for the two-dimensional BC model
in (a) Critical point (b) Tricritical point. These plots exactly ilustrate
which is theoretically predicted in \protect\cite{Janssen1989} and 
\protect\cite{Janssen1994} respectively. }
\label{Fig:critical_and_tricritical_2D}
\end{figure}

Fig. \ref{Fig:critical_and_tricritical_2D} shows exactly these time
evolutions which corroborates the Refs. \cite{Janssen1989} and \cite%
{Janssen1994}. This same figure corroborates that when $m_{0}\rightarrow 1$,
i..e, from ordered initial systems, one expects a decay: 
\begin{equation}
m_{m_{0}=1}(t)\sim t^{-\lambda }\text{.}  \label{Eq:ord}
\end{equation}%
where $\lambda =\frac{\beta }{\nu z}$. Here $\beta $ is related to
magnetization, and $\nu $ related to correlation length $\xi $, and a
dynamic exponent $z$ which links the length correlation with time
correlation $\Delta $ such that $z\sim \frac{\ln \Delta }{\ln \xi }$, is the
dynamic exponent.

The exponent $z$ can be extracted, in a independent way, from the ratio \cite%
{SilvaPLA2002}: 
\begin{equation}
F_{2}(t)=\frac{m_{m_{0}=0}^{(2)}(t)}{m_{m_{0}=1}^{2}(t)}\sim t^{\varsigma }
\label{Eq:ratio}
\end{equation}%
where $\varsigma =d/z$, since $m_{m_{0}=0}^{(2)}(t)=\frac{1}{N^{2}}%
\left\langle \left( \sum_{i=1}^{N}\sigma _{i}(t)\right) ^{2}\right\rangle
\sim t^{(d-2\beta /\nu )/z}$, where $m_{0}=0$, but with spins randomly
distributed. With exponent $z$, one estimates $\nu $ from a power law
expected by the derivative of log-magnetization \cite{Zheng1998}: 
\begin{equation}
D(t)=\frac{1}{2\delta }\ln \left[ \frac{m_{m_{0}=1}(T_{c}+\delta )}{%
m_{m_{0}=1}(T_{c}-\delta )}\right] \sim t^{\vartheta }  \label{Eq:derivada}
\end{equation}%
with $\vartheta =\frac{1}{\nu z}$ which is valid when $\delta <<1$, and
finally with $\nu $ and $z$ in hands, we return to \ref{Eq:ord} and
calculate $\beta $. These three power laws can be used to determine $\beta
,\ \nu $ and $z$ for the critical (see for example \cite{Critical}) and
tricritical \cite{SilvaPRE2002,SilvaTRICPC2013} in both two and
three-dimensional nonequilibrium simulations which is known as short-time
dynamics, but the question is: could they be observed in mean-field regime?
Moreover, how works the initial anomalous behavior of the magnetization for $%
m_{0}<<1$?

In this paper, we will show the peculiarities of the nonequilibrium critical
dynamics of mean-field Ising-like systems for both: spin 1/2 (Ising) and
spin 1 (BC). We will localize the critical temperature with a method
previously used for systems with SR interactions \cite{SilvaPRE2012} but
that we here show to be properly interesting to estimate the critical
parameters in MF systems. After that, we show that there is a notorious
difference between the relaxation of the nonequilibrium critical dynamics
from disordered initial systems for two-dimensional and MF systems, via MC
simulations and some analytical results. However, we show that power laws
are described by Eqs. \ref{Eq:ord}, \ref{Eq:ratio}, and \ref{Eq:derivada}
are valid in the MF regime and the exponents $\beta ,\ \nu $ and $z$ were
calculated corroborating the classical exponents in both: critical and
tricritical points. Finally, the persistence phenomenon is also investigated
in the MF regime. In section \ref{Sec:MF-Equations} we develop the equations
that describe the time evolution of the magnetization of Ising and BC
models, valid for critical and tricritical points as a function of the
initial magnetization. Our results are presented in section \ref{Sec:Results}%
. Our main conclusions are resumed in section \ref{Sec:Conclusions}.

\section{Mean-Field MC dynamics for Ising-like systems and effects of
initial magnetization}

\label{Sec:MF-Equations}

In discrete systems, as Ising-like ones, if one denotes $\Pr (\sigma ,t)$ is
the probability of the occurrence of configuration $\sigma =\left( \sigma
_{1},...,\sigma _{N}\right) $ at time $t$. Thus, the expected value of a
spin at site $j=1,...,\ N$, is given by%
\begin{equation}
\left\langle \sigma _{j}\right\rangle =\sum_{\sigma }\sigma _{j}\Pr (\sigma
,t)
\end{equation}

By using the master equation, and for systems with spin 1/2 (the usual Ising
model), one can show as direct consequence of the master equation (see for
example \cite{MarioBook}) that:%
\begin{equation*}
\frac{d\left\langle \sigma _{j}\right\rangle }{dt}=-2\sum_{\sigma }\sigma
_{j}w(\sigma \rightarrow \sigma ^{(j)})\Pr (\sigma ,t)\text{,}
\end{equation*}%
where $w(\sigma \rightarrow \sigma ^{(j)})$ is the transition rate to from $%
\sigma $ to $\sigma ^{(j)}$, where this latter denotes $\sigma ^{(i)}=\left(
\sigma _{1},...\sigma _{i-1},-\sigma _{i},\sigma _{i+1},\sigma _{N}\right) $
which only difference is that $\sigma _{i}$ is substituted by $-\sigma _{i}$%
, which represents a local transition in the regular representation.

We can choose any prescription that satisfies the detailed balance, for
example the usual Glauber dynamics:

\begin{equation*}
w(\sigma \rightarrow \sigma ^{(j)})=\frac{1}{\tau }\frac{e^{-\beta \Delta 
\mathcal{H}_{MF}}}{1+e^{-\beta \Delta \mathcal{H}_{MF}}}
\end{equation*}%
where $\tau $ is a characteristic parameter that can be made equal to 1, or
fitted to contemplate the time scale of a desired numerical simulation. It
is most often found as:

\begin{equation*}
w(\sigma \rightarrow \sigma ^{(j)})=\frac{1}{2\tau }\left[ 1-\tanh \left(
\beta \frac{\Delta \mathcal{H}_{MF}}{2}\right) \right]
\end{equation*}

In the case of the mean-field Ising model, one has 
\begin{equation*}
\begin{array}{lll}
\Delta \mathcal{H}_{MF} & = & -\frac{J}{2N}z\left[ \left( \sum_{i=1,\ i\neq
j}^{N}\sigma _{i}-\sigma _{j}\right) ^{2}\right. \\ 
&  &  \\ 
&  & \left. -\left( \sum_{i=1,\ i\neq j}^{N}\sigma _{i}+\sigma _{j}\right)
^{2}\right] \\ 
&  &  \\ 
& = & \frac{2J}{N}z\sigma _{j}\sum_{i=1,\ i\neq j}^{N}\sigma _{i} \\ 
&  &  \\ 
& = & \frac{2J}{N}z\sigma _{j}\left( M-\sigma _{j}\right) \approx \frac{2J}{N%
}z\sigma _{j}M%
\end{array}%
\end{equation*}%
Thus, $w(\sigma \rightarrow \sigma ^{(j)})=\frac{1}{2\tau }\left[ 1-\tanh
\left( \frac{\beta J}{N}z\sigma _{j}M\right) \right] $, and since $\tanh (x)$
is a odd function, $w(\sigma \rightarrow \sigma ^{(j)})=\frac{1}{2\tau }%
\left[ 1-\sigma _{j}\tanh \left( \frac{\beta J}{N}zM\right) \right] $.

Thus, since $\sigma _{j}^{2}=1$ 
\begin{equation*}
\begin{array}{lll}
\frac{d\left\langle \sigma _{j}\right\rangle }{dt} & = & -\frac{1}{\tau }%
\left[ \sum_{\sigma }\sigma _{j}\Pr (\sigma ,t)\right. \\ 
&  &  \\ 
&  & \left. -\sum_{\sigma }\sigma _{j}^{2}\tanh \left( \frac{\beta J}{N}%
zM\right) \Pr (\sigma ,t)\right] \\ 
&  &  \\ 
& = & -\frac{1}{\tau }\left[ \left\langle \sigma _{j}\right\rangle
-\left\langle \tanh \left( \frac{\beta J}{N}zM\right) \right\rangle \right]%
\end{array}%
\end{equation*}

Since $\left\langle \sigma _{j}\right\rangle =m$, we have and considering
that in the mean-field approximation one has$\ \left\langle \tanh \left( 
\frac{\beta J}{N}zM\right) \right\rangle =\tanh \left\langle \frac{\beta J}{N%
}zM\right\rangle =\tanh \beta zJm$, one obtains: 
\begin{equation}
\tau \frac{dm}{dt}=-m+\tanh \beta Jzm
\end{equation}%
where $m=\lim_{N\rightarrow \infty }\frac{\left\langle M\right\rangle }{N}$
since one has $N>>1$.

It is important to observe that the right side of this equation is exactly
the negative of free energy of the Ising model: 
\begin{equation*}
\tau \frac{dm}{dt}=\left. \frac{\partial f}{\partial y}\right\vert _{y=m}
\end{equation*}%
where $f(y)=\frac{\Phi (y,h=0)}{Jz}=\frac{y^{2}}{2}-\frac{1}{\beta Jz}\ln
(2\cosh (\beta Jzy))$. Sure for $\beta Jz=1$ (critical point), $m<<1$, and
thus $\tau \frac{dm}{dt}=-m+\tanh m\approx -\frac{1}{3}m^{3}$. And then

\begin{equation}
m(t)=m_{0}\sqrt{\frac{3}{3+2m_{0}^{2}t/\tau }}\sim t^{-1/2}
\label{Eq:Initial_magnetization_behavior}
\end{equation}%
for $t\rightarrow \infty $.

Taking this idea and transposing it for the Blume Capel Model: 
\begin{equation}
f(y)=\frac{y^{2}}{2}-\frac{1}{\beta Jz}\ln (2e^{-\beta D}\cosh (\beta Jzy)+1)
\label{Eq:energia_livre_BC}
\end{equation}%
one has 
\begin{equation}
\tau \frac{dm}{dt}=-m+\frac{2e^{-\beta D}\sinh (\beta Jzm)}{2e^{-\beta
D}\cosh (\beta Jzm)+1}  \label{Eq:critical_dynamics_MF_BC}
\end{equation}

The critical line in mean field regime is given by:$\frac{D}{Jz}=\frac{K_{B}T%
}{Jz}\ln \left[ \frac{2(Jz-K_{B}T)}{K_{B}T}\right] $, thus $\frac{D}{K_{B}T}%
=\beta D=\ln \left[ 2(\beta Jz-1)\right] =\ln (2(\alpha -1))$, where $\alpha
=\beta Jz$. A special point is the tricritical one:$\frac{D}{K_{B}T_{t}}%
=2\ln 2 $, which leads to $\alpha _{t}=\frac{Jz}{k_{B}T_{t}}=3$. In this
paper one considers only the critical line: from $\frac{D}{Jz}=0$ until the
critical point $\frac{D}{Jz}=\frac{2}{3}\ln 2$. For $\frac{D}{Jz}>$ $\frac{2%
}{3}\ln 2$ one has first order transition points.

Thus one has: $\tau \frac{dm}{dt}=-m+\frac{(\alpha -1)^{-1}\sinh (\alpha m)}{%
(\alpha -1)^{-1}\cosh (\alpha m)+1}$, and considering the approximations: $%
\sinh (\alpha m)=\allowbreak m\alpha +\frac{1}{6}m^{3}\alpha ^{3}+\frac{1}{%
120}m^{5}\alpha ^{5}+O\left( m^{7}\right) $ and $\cosh (\alpha
m)=\allowbreak 1+\frac{1}{2}m^{2}\alpha ^{2}+\frac{1}{24}m^{4}\alpha
^{4}+O\left( m^{6}\right) $, one has executing few steps of algebra that:

\begin{equation}
\begin{array}{lll}
\tau \frac{dm}{dt} & = & -m+\frac{(\alpha -1)^{-1}[m\alpha +\frac{1}{6}%
m^{3}\alpha ^{3}+\frac{1}{120}m^{5}\alpha ^{5}+O\left( m^{7}\right) ]}{%
(\alpha -1)^{-1}[\allowbreak 1+\frac{1}{2}m^{2}\alpha ^{2}+\frac{1}{24}%
m^{4}\alpha ^{4}+O\left( m^{6}\right) ]+1} \\ 
&  &  \\ 
& = & -m+\frac{m\left[ 1+\frac{1}{6}m^{2}\alpha ^{2}+\frac{1}{120}%
m^{4}\alpha ^{4}+O\left( m^{6}\right) \right] }{\left[ 1+\frac{1}{2}%
m^{2}\alpha +\frac{1}{24}m^{4}\alpha ^{3}+O\left( m^{6}\right) \right] } \\ 
&  &  \\ 
& = & -m^{3}\left( \frac{1}{2}\alpha -\frac{1}{6}\alpha ^{2}\right) + \\ 
&  &  \\ 
&  & m^{5}\left( \frac{1}{120}\alpha ^{4}-\frac{1}{8}\alpha ^{3}+\frac{1}{4}%
\alpha ^{2}\right) +\allowbreak O\left( m^{7}\right) \\ 
&  &  \\ 
& \approx & c_{3}m^{3}+c_{5}m^{5}%
\end{array}
\label{Eq:differential_equation_MF_BC}
\end{equation}%
where 
\begin{equation}
c_{3}=\frac{1}{6}\alpha ^{2}-\frac{1}{2}\alpha \text{ }  \label{Eq:c3}
\end{equation}%
and

\begin{equation}
c_{5}=\frac{1}{120}\alpha ^{4}-\frac{1}{8}\alpha ^{3}+\frac{1}{4}\alpha ^{2}
\label{Eq:c5}
\end{equation}%
since at the critical (or tricritical) points $m\approx 0$. The general
solution is then given by:

\begin{eqnarray}
k+\frac{t}{\tau } &=&-\frac{c_{5}}{c_{3}^{2}}\ln m(t)+
\label{Eq:transcedental} \\
&&\frac{c_{5}}{2c_{3}^{2}}\ln (c_{3}+c_{5}m^{2}(t))-\frac{1}{2c_{3}m^{2}(t)}
\end{eqnarray}%
where $k=\frac{c_{5}}{2c_{3}^{2}}\ln \frac{\sqrt{(c_{3}+c_{5}m_{0}^{2})}}{%
m_{0}}-\frac{1}{2c_{3}m_{0}^{2}}$, which is a transcendental equation. More
details of this solution will be discussed in the next section.

\section{Results}

\label{Sec:Results}

We perform TDMC simulations in the MF regime. We start with the standard
Ising model (spin 1/2). We prepare $N$ spins such that the total initial
magnetization is $M_{0}=\sum_{i=1}^{N}\sigma _{i}(0)$ (in our notation $%
m_{0}=\frac{M_{0}}{N}$ is the initial magnetization per spin). Thus, at
temperature $T$, we evolve the system by $N_{steps}$ MC steps, under $%
N_{run} $ different runs. Each $t\ $th MC step considers that $N$ spins are
drawn and each drawn spin $\sigma _{i}$ is flipped, for example, with
MF-Metropolis probability:

\begin{equation}
\begin{array}{lll}
p(\sigma _{i}\rightarrow -\sigma _{i}) & = & \min \{1,\exp (-\beta \Delta 
\mathcal{H}_{MF})\} \\ 
&  &  \\ 
& = & \min \left\{ 1,\exp \left[ \frac{2Jz}{Nk_{B}T}(1-\sigma _{i}M(t))%
\right] \right\}%
\end{array}
\label{Eq:Metropolis_Ising}
\end{equation}%
where $M(t)=\sum_{i=1}^{N}\sigma _{i}(t)\ $is the magnetization at time $t$.
The simulated moments are calculated by: $m_{m_{0}}^{(k)}(t)=\frac{1}{N^{k}}%
\left\langle \left( \sum_{i=1}^{N}\sigma _{i}(t)\right) ^{k}\right\rangle
_{MC}=\frac{1}{N_{run}N^{k}}\sum_{j=1}^{N_{run}}\left( \sum_{i=1}^{N}\sigma
_{i,j}(t)\right) ^{k}$, where $\sigma _{i,j}(t)$ denotes the $i$-th spin, at 
$j$-th run, at $t-$th time step.

\begin{figure}[tbp]
\begin{center}
\includegraphics[width=1.0\columnwidth]{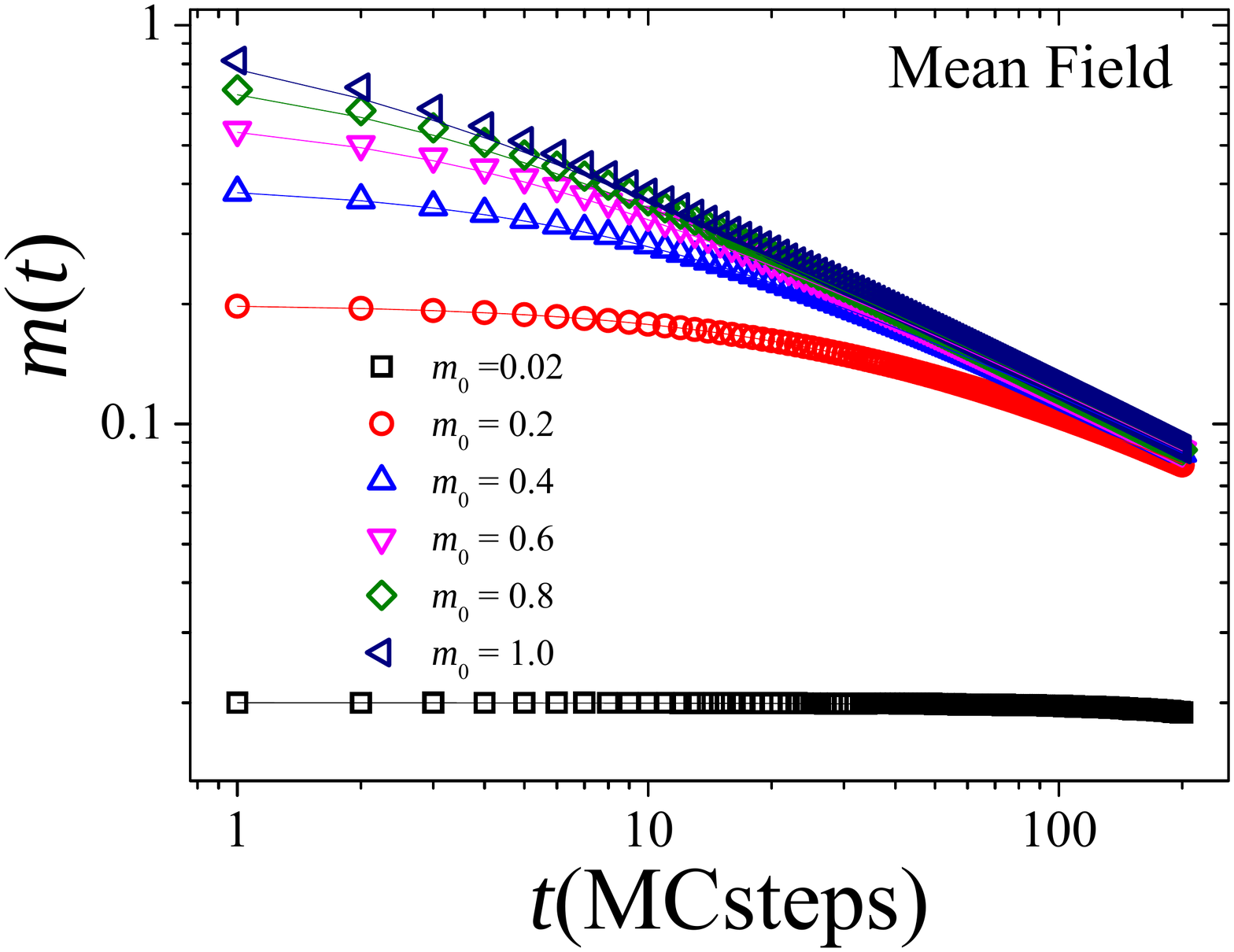}
\end{center}
\caption{Time evolution of magnetization for different values of $m_{0}$ of
the MF Ising model. Points correspond to MC simulations while curves
correspond to fit with the function from Eq. \protect\ref%
{Eq:Initial_magnetization_behavior}.}
\label{Fig:time_evolution_ising_different_m0}
\end{figure}

Fig. \ref{Fig:time_evolution_ising_different_m0} shows the time evolution of
the magnetization for different values of $m_{0}$ for $\frac{k_{B}T_{c}}{Jz}%
=1$. Points corresponds to MC simulations while curves corresponds to fit
with function of the Eq. \ref{Eq:Initial_magnetization_behavior}. We have a
good agreement between simulation and theory. We have calibrated $\tau $ to
perform the fits for each different value of $m_{0}$. For such simulations,
we used $N=10^{5}$ spins, $N_{run}=15000$ runs. Different from MC
simulations for two or three-dimensional lattices, we do not observe an
initial slip of the magnetization as predicted by mean-field equations.

For $m_{0}=1$, for $t>t_{\min }\approx 10$ MC steps, one observes the
expected power-law $m(t)\sim t^{-\lambda }$, where $\lambda $ is expected to
be equal to 1/2 from theory. In order to check if the critical temperature
is correct one can use the method developed in \cite{SilvaPRE2012} and
highly used in several different models with \cite{With} and without \cite%
{Without} defined Hamiltonian, which find the optimal $K=\frac{k_{B}T}{Jz}$,
denoted by $K^{(opt)}$, for which $m(t)\times t$ leads to the best power
law. The idea is simple since at criticality it is expected that the order
parameter obeys the power-law behavior given by $m(t)\sim t^{-\lambda }$, we
performed MC simulations for each value $K=K^{(\min )}+i\Delta K$, with $%
i=1,...,n$, where $n=\left\lfloor (K^{(\max )}-K^{(\min )})/\Delta
K\right\rfloor $, and calculated the coefficient of determination $r$, which
is given by 
\begin{equation}
r=\frac{\sum\limits_{t=t_{\min }}^{t_{\max }}(\overline{\ln m}-a-b\ln t)^{2}%
}{\sum\limits_{t=t_{\min }}^{t_{\max }}(\overline{\ln m}-\ln m(t))^{2}},
\label{determination_coefficient}
\end{equation}%
with $\overline{\ln m}=\frac{1}{(t_{\max }-t_{\min })}\sum\nolimits_{t=t_{%
\min }}^{t_{\max }}\ln m(t)$. The critical value $K_{c}=\frac{k_{B}T_{c}}{Jz}
$ corresponds to $K^{(opt)}=\arg \max_{K\in \lbrack K^{(\min )},K^{(\max
)}]}\{r\}$ and, $a$ and $b$ are, respectively, the slope and intercept
obtained from the linearization. Here, $t_{\min }$ is the number of
discarded MC steps and $t_{\max }$ the maximum number of MC steps used in
our simulations.

\begin{figure}[tbp]
\begin{center}
\includegraphics[width=1.0\columnwidth]{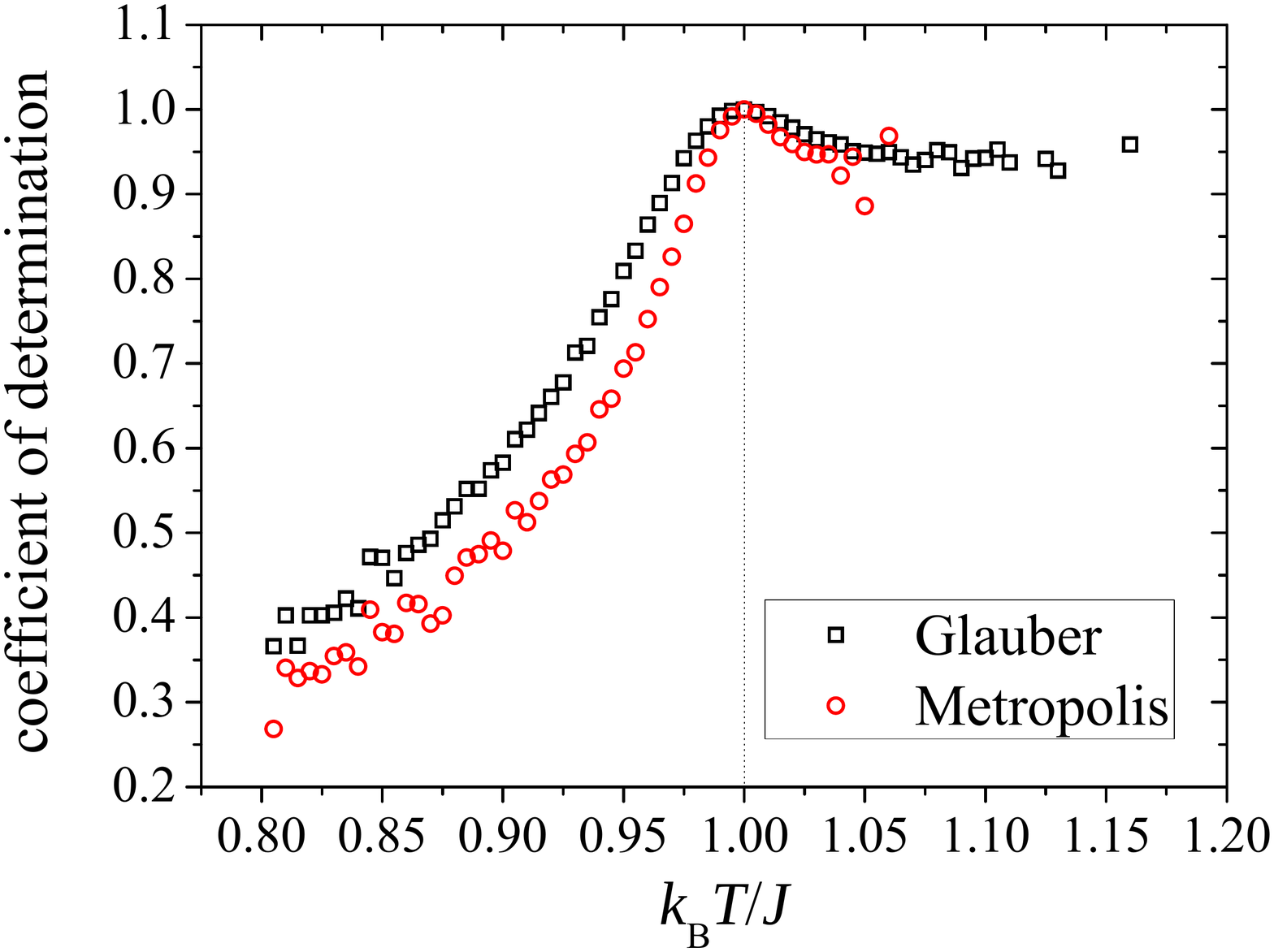} %
\includegraphics[width=1.0\columnwidth]{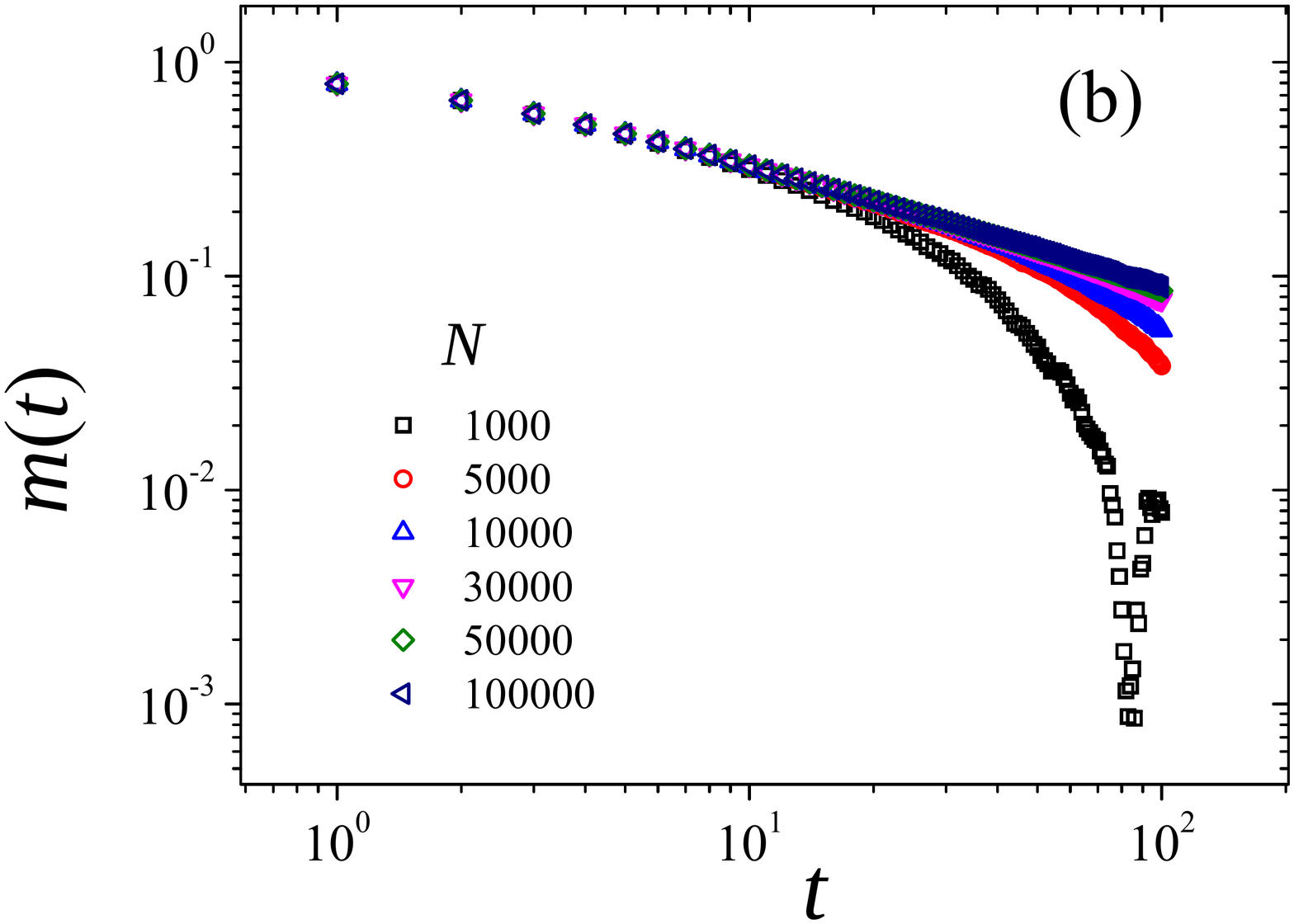}
\end{center}
\caption{(a) Determination coefficient as a function of temperature. The
optimal value is obtained exactly in the critical temperature ($K_{c}=\frac{%
k_{B}T_{c}}{Jz}\approx 1$). (b) Finite-size study of decay of magnetization.
We can observe robust power laws for $L=10^{5}$ spins. We adopted such a
size for our simulations in this paper. }
\label{Fig:Localization_and_fss}
\end{figure}

Fig. \ref{Fig:Localization_and_fss} (a) shows the coefficient of
determination as function of $K$. We obtain curves for both Metropolis (Eq. %
\ref{Eq:Metropolis_Ising}) and we also added results for Glauber dynamics: $%
p(\sigma _{j}\rightarrow -\sigma _{j})=\frac{1}{2}\left[ 1-\tanh \left( 
\frac{Jz}{Nk_{B}T}\sigma _{j}M(t)\right) \right] $. We observe that
regardless of the dynamics, the maximum value of $r$ is obtained for $%
K_{c}\approx 1$, corroborating that TDMC in the MF regime can determine the
critical temperature of the spin-1/2 Ising model.

To avoid doubts about the required size of systems to be used, fig. \ref%
{Fig:Localization_and_fss} (b) shows a finite-size study observing the
magnetization decay for different sizes. We can observe that a robust power
law is obtained for $L=10^{5}$. In this paper, we used $L=10^{5}$ spins to
obtain all results.

Now, it is interesting to observe if the critical exponents of the MF Ising
model are numerically corroborated. In mean-field is expected $\nu =1/2$, $%
\beta =1/2$, and $z=2$ \cite{MarioBook}, which means $\lambda =1/2$, $%
\vartheta =2$. For $\varsigma $ that for SR system corresponds to $d/z$, in
MF regime it is a little more confused. There is a rule which says that
above a certain critical dimension (upper critical dimension) the exponents
are given by the MF theory (classical exponents). Therefore, when we have a
scaling relation where $d$ is considered, we must use the critical dimension
that in the case of Ising model is $d_{c}=4$ \cite{MarioBook,Pleimling}.
Therefore we must do $\varsigma =4/z$ in the MF regime, which results in $%
\varsigma =2$ for the Ising model. However for the tricritical point $%
d_{c}=3 $ \cite{Lawrie}, and then in this case: $\varsigma =3/z$.

\begin{figure}[tbp]
\begin{center}
\includegraphics[width=1.0\columnwidth]{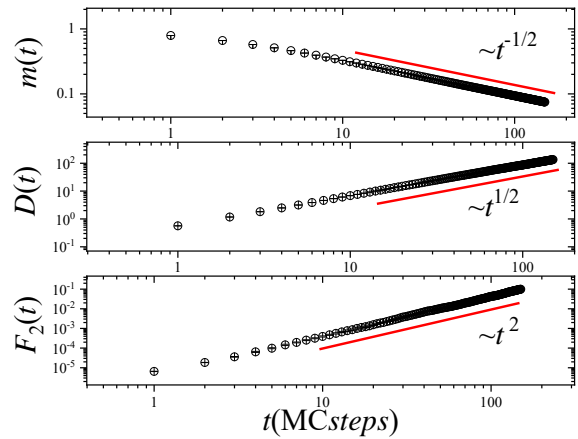}
\end{center}
\caption{Different power laws, Eqs. \protect\ref{Eq:ord}, \protect\ref%
{Eq:ratio}, and \protect\ref{Eq:derivada}, for $K_{c}=\frac{k_{B}T_{c}}{Jz}%
=1 $. }
\label{Fig:power_laws_critical_exponent}
\end{figure}

Fig. \ref{Fig:power_laws_critical_exponent} shows the simulated power laws
corroborating the ones from Eqs. \ref{Eq:ord}, \ref{Eq:ratio}, and \ref%
{Eq:derivada} for the MF Ising model.

Since these power laws were obtained for $m_{0}=1$, $N_{run}=2000$ runs
which are more than enough in this condition. The uncertainty bars are
obtained performing $N_{bin}=5$ five different sets corresponding to
different seeds in our simulations. The results lead to: $\lambda =0.5484(6)$%
, $\varsigma =2.033(3)$, and $\vartheta =1.078(2)$. This leads to $\beta
=0.509\,(1)$, $z=1.\,\allowbreak 967\,(3)$, and $\nu =0.472\,(1)$ that
corroborate the critical classical exponents.

Since we have performed a preliminary study with the Ising model, now we can
explore the critical-tricritical behavior of the MF Blume Capel model. In
this case, for the sake of simplicity, here one uses heat-bath dynamics.
According to Eq. \ref{Eq:long_hange_hamiltonian} one has that energy in a BC
model ($H=0$) with current spin $\sigma _{i}$ replaced by spin $\sigma
_{i}^{\prime }$ is given by: 
\begin{equation*}
\begin{array}{lll}
E(\sigma _{i}^{\prime }) & = & -\frac{Jz}{2N}(M-\sigma _{i}+\sigma
_{i}^{\prime })^{2} \\ 
&  &  \\ 
&  & +\ D\sum\limits_{\substack{ k=1  \\ k\neq i}}^{N}\sigma
_{k}^{2}+D\sigma _{i}^{\prime 2}%
\end{array}%
\end{equation*}%
thus the transition probability to spin $\sigma _{i}^{\prime }$ is:%
\begin{equation*}
p(\sigma _{i}^{\prime })=\frac{e^{-\beta E(\sigma _{i}^{\prime })}}{%
e^{-\beta E(0)}+e^{-\beta E(+)}+e^{-\beta E(-)}}
\end{equation*}%
and thus: $p(0)=\left[ 1+e^{-\beta (E(+)-E(0))}+e^{-\beta (E(-)-E(0))}\right]
^{-1}$, similarly $p(-)=\left[ 1+e^{-\beta (E(+)-E(-))}+e^{-\beta
(E(0)-E(-))}\right] ^{-1}$, and naturally $p(+)=1-p(0)-p(-)=\left[
1+e^{-\beta (E(-)-E(+))}+e^{-\beta (E(0)-E(+))}\right] ^{-1}$.

Thus what we have calculate is $\Delta E_{x,y}=E(x)-E(y)$, which is given by:%
\begin{equation}
\begin{array}{lll}
\Delta E_{x,y} & = & -\frac{Jz}{2N}\left[ (M-\sigma _{i}+x)^{2}-(M-\sigma
_{i}+y)^{2}\right]  \\ 
&  & +\ D(x^{2}-y^{2}) \\ 
&  &  \\ 
& = & -\frac{Jz}{2N}\left[ 2\left( M-\sigma _{i}\right) (x-y)\right]  \\ 
&  & +\left( D-\frac{Jz}{2N}\right) (x^{2}-y^{2}) \\ 
&  &  \\ 
& \approx  & -Jz\frac{M}{N}(x-y)+D(x^{2}-y^{2})%
\end{array}
\label{Eq:variation}
\end{equation}

However, to avoid any doubt, one used the exact form of equation (second
line in the \ref{Eq:variation}) in our simulations, and not the
approximation suggested in the last line of this same equation.

Thus, by fixing the value of $D/k_{B}T$, we change $K=\frac{k_{B}T}{Jz}$ to
obtain $K_{c}$, i.e., the value that maximizes $r$. One uses $\Delta K=0.005$%
. The values used are $D/k_{B}T=0$, $0.28$, $0.54$, $0.84$, $1.12$, and $\ln
4$, this last one corresponding to the tricritical point. Fig \ref%
{Fig:critical_tricritical_localization} (a) shows the coefficient of
determination as function of $K=\frac{k_{B}T}{Jz}$. We can observe that
critical temperatures, corresponding to the maximum values of coefficient of
determination, are found in all studied cases. 
\begin{figure}[tbp]
\begin{center}
\includegraphics[width=1.0\columnwidth]{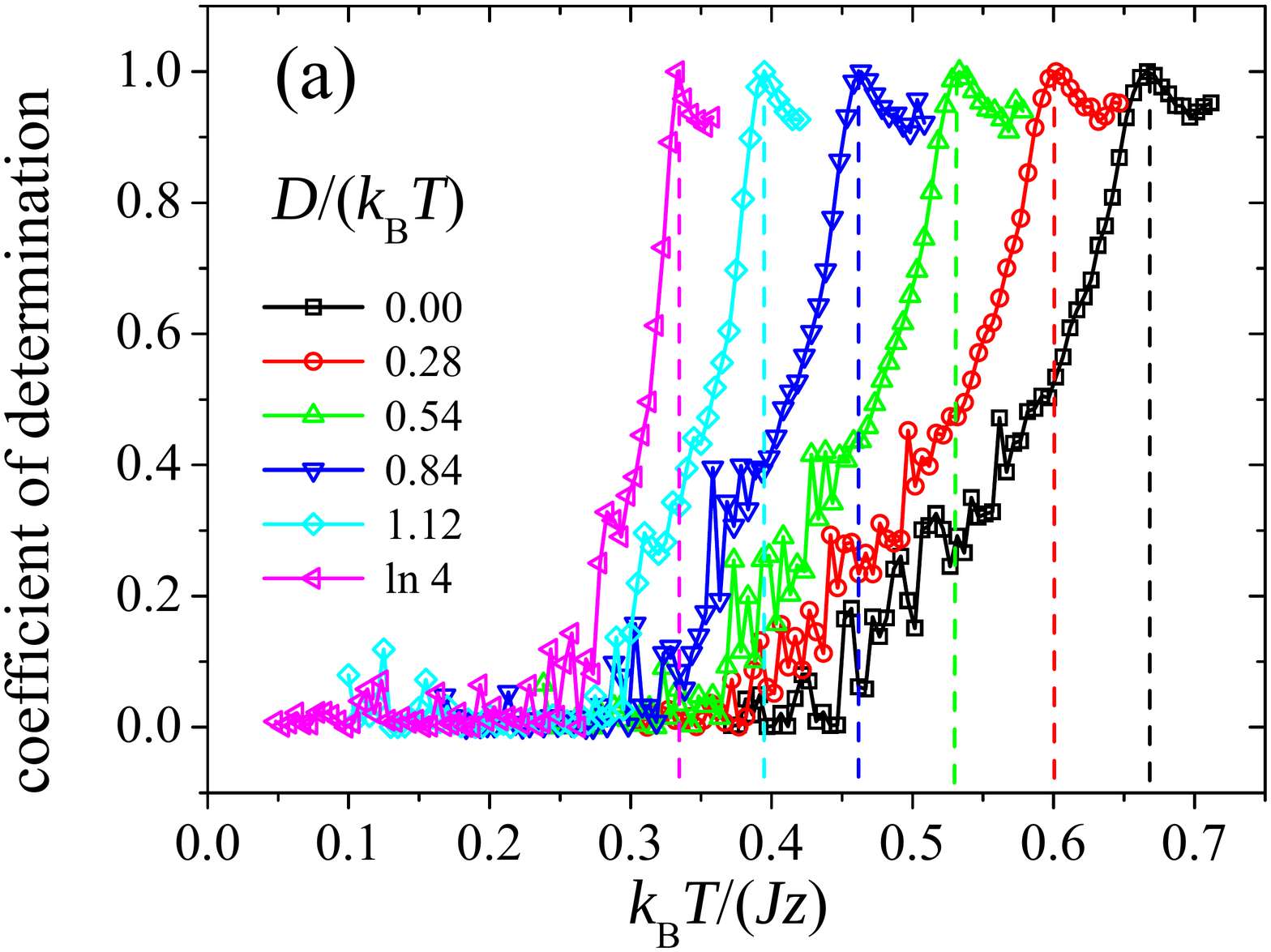} %
\includegraphics[width=1.0\columnwidth]{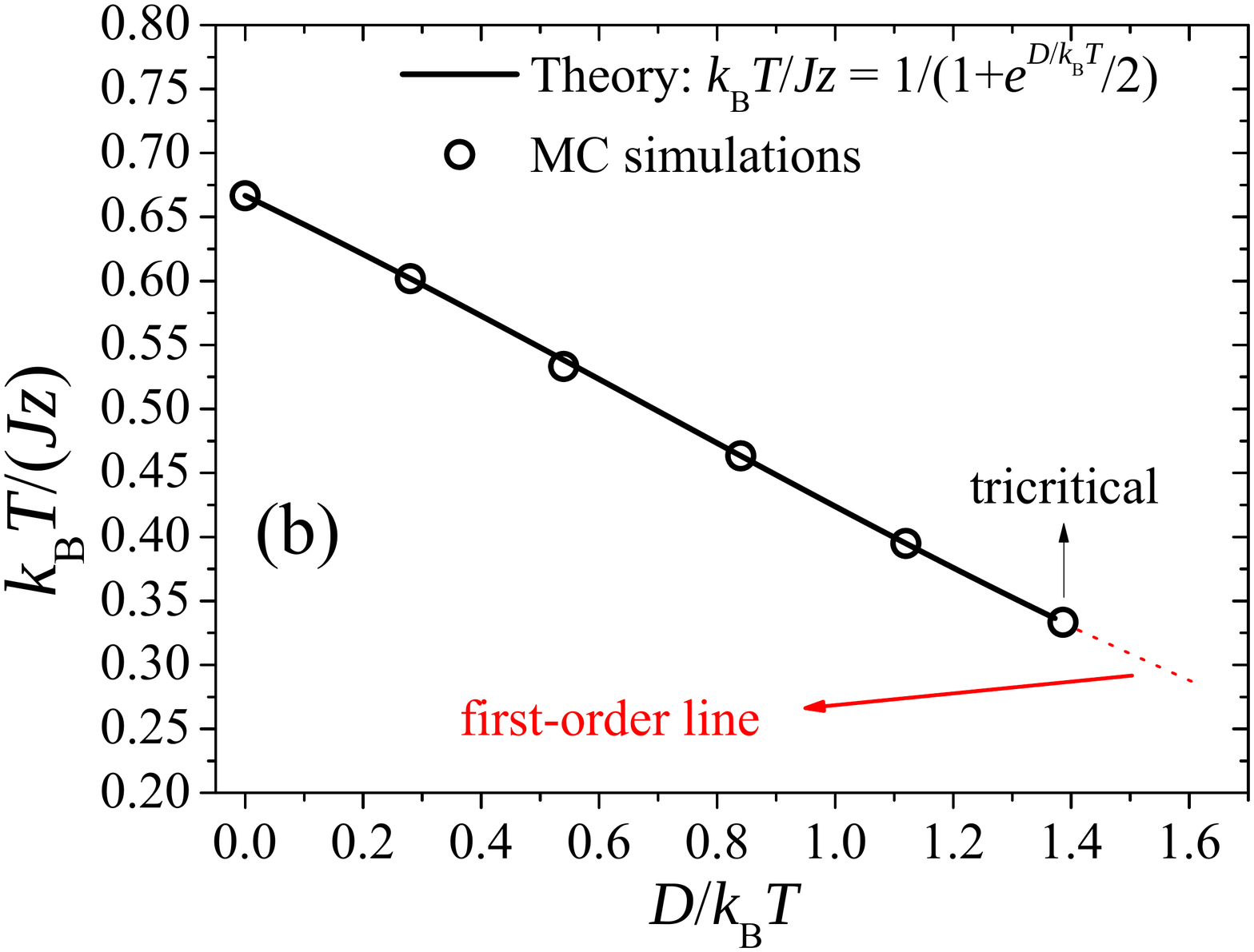}
\end{center}
\caption{(a) Coefficient of determination for different values of $D/k_{B}T$%
. (b) Critical line: comparison between theory and TDMC simulations in the
MF regime. }
\label{Fig:critical_tricritical_localization}
\end{figure}
The corresponding Fig. \ref{Fig:critical_tricritical_localization} (b) is
build with values found in Fig \ref{Fig:critical_tricritical_localization}
(a). We can observe an excellent agreement between the results found in our
simulations with theoretical prediction, which shows that TDMC simulations
in the MF regime can be performed as occurs in regular time-dependent MC
simulations in lattices.

Since one has $\tau \frac{dm}{dt}=c_{3}m^{3}+c_{5}m^{5}$, it is interesting
to check two extremal cases: $c_{3}=0$, and $c_{5}=0$. For example, when one
has $c_{3}=0$, it leads to $\frac{1}{6}\alpha ^{2}-\frac{1}{2}\alpha =0$
that has as non-trivial solution $\alpha _{t}=\frac{Jz}{k_{B}T_{t}}=3$. This
corresponds to $\tau \frac{dm}{dt}=-\frac{9}{20}m^{5}$, which solution is 
\begin{equation}
m_{\text{tri}}(t)=\frac{1}{(m_{0}^{-4}+\frac{9}{5\tau }t)^{1/4}}
\end{equation}%
which results in%
\begin{equation}
m_{\text{tri}}(t)\sim t^{-1/4}  \label{Eq:tricritical_power_law}
\end{equation}%
for $t\rightarrow \infty $. On the other hand, if $c_{5}=\frac{1}{4}\alpha
^{2}(\frac{\alpha ^{2}}{30}-\frac{\alpha }{2}+1)=0$, it has as non-trivial
solution $\alpha _{c}^{(1)}=\frac{15}{2}-\frac{1}{2}\sqrt{105}$ and $\alpha
_{c}^{(2)}=\frac{15}{2}+\frac{1}{2}\sqrt{105}$, but $\alpha _{c}^{(2)}=\frac{%
15}{2}+\frac{1}{2}\sqrt{105}>3=\alpha _{\text{tri}}$, which belongs to first
order region. Thus we must concentrate our attentions to $\alpha _{c}^{(1)}=%
\frac{15}{2}-\frac{1}{2}\sqrt{105}\approx \allowbreak 2.\,\allowbreak
376\,52 $. Thus $\tau \frac{dm}{dt}=\left( -\sqrt{105}+10\right) m^{3}$, which
leads to%
\begin{equation*}
m_{\text{cri}}(t)=\frac{1}{\sqrt{m_{0}^{-2}+(2/\tau )(\sqrt{105}-10)t}}.
\end{equation*}%
Asymptotically one has $m_{\text{cri}}(t)\sim t^{-1/2}$ that corresponds to
the behavior of the only acceptable critical point: $\alpha _{c}^{(1)}=\frac{%
15}{2}-\frac{1}{2}\sqrt{105}$. However, when we perform TDMC simulations in
the MF regime, one should observe such similar power law to other critical
points but a crossover to the tricritical point is also expected since
according to Eq. \ref{Eq:tricritical_power_law} the exponent changes.

\begin{figure}[tbp]
\begin{center}
\includegraphics[width=1.0\columnwidth]{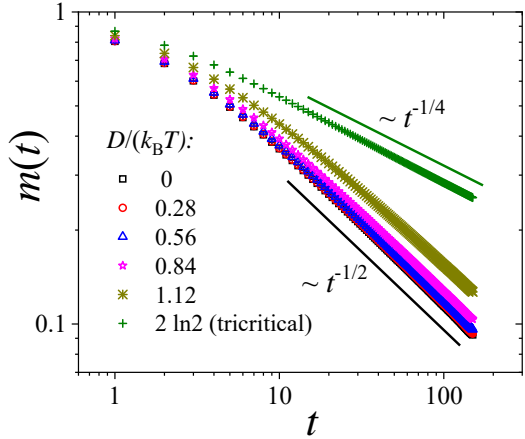}
\end{center}
\caption{Decay of the magnetization for different values of $D/k_{B}T$:$\ 0$%
, $0.28$, $0.56$, $0.84$, $1.12$, and $2\ln 2$ (tricritical). }
\label{Fig:Decay_mag_BC}
\end{figure}

Thus, we start by studying $m(t)$, starting from $m_{0}=1$, for all points
which we localized according to Fig. \ref%
{Fig:critical_tricritical_localization}. The idea is to study how $\lambda $
changes as function of $D/(k_{B}T)$. Fig. \ref{Fig:Decay_mag_BC} shows $m(t)$
as function of $t$ for the different values of $D/(k_{B}T)$.

\begin{table*}[tbp] \centering%
\begin{tabular}{lllllll}
\hline\hline
$\frac{D}{k_{B}T}$ & $0.00$ & $0.28$ & $0.56$ & $0.84$ & $1.12$ & $2\ln 2$
(TP) \\ \hline\hline
$\lambda $ & 0.5162(4) & 0.5119(3) & 0.5096(3) & 0.5014(3) & 0.4724(5) & 
0.2665(1) \\ 
$\varsigma $ & 1.951(2) & 1.954(3) & 1.950(3) & 1.952(2) & 1.897(2) & 
1.529(1) \\ 
$\vartheta $ & 1.0362(6) & 1.0333(5) & 1.0299(6) & 1.0242(3) & 1.0141(3) & 
1.0258(2) \\ \hline\hline
\end{tabular}%
\caption{Mean-field raw exponents from Blume Capel model along the critical line
obtained with Monte Carlo simulations}\label{Table:raw}%
\end{table*}%

\begin{figure}[tbp]
\begin{center}
\includegraphics[width=1.0\columnwidth]{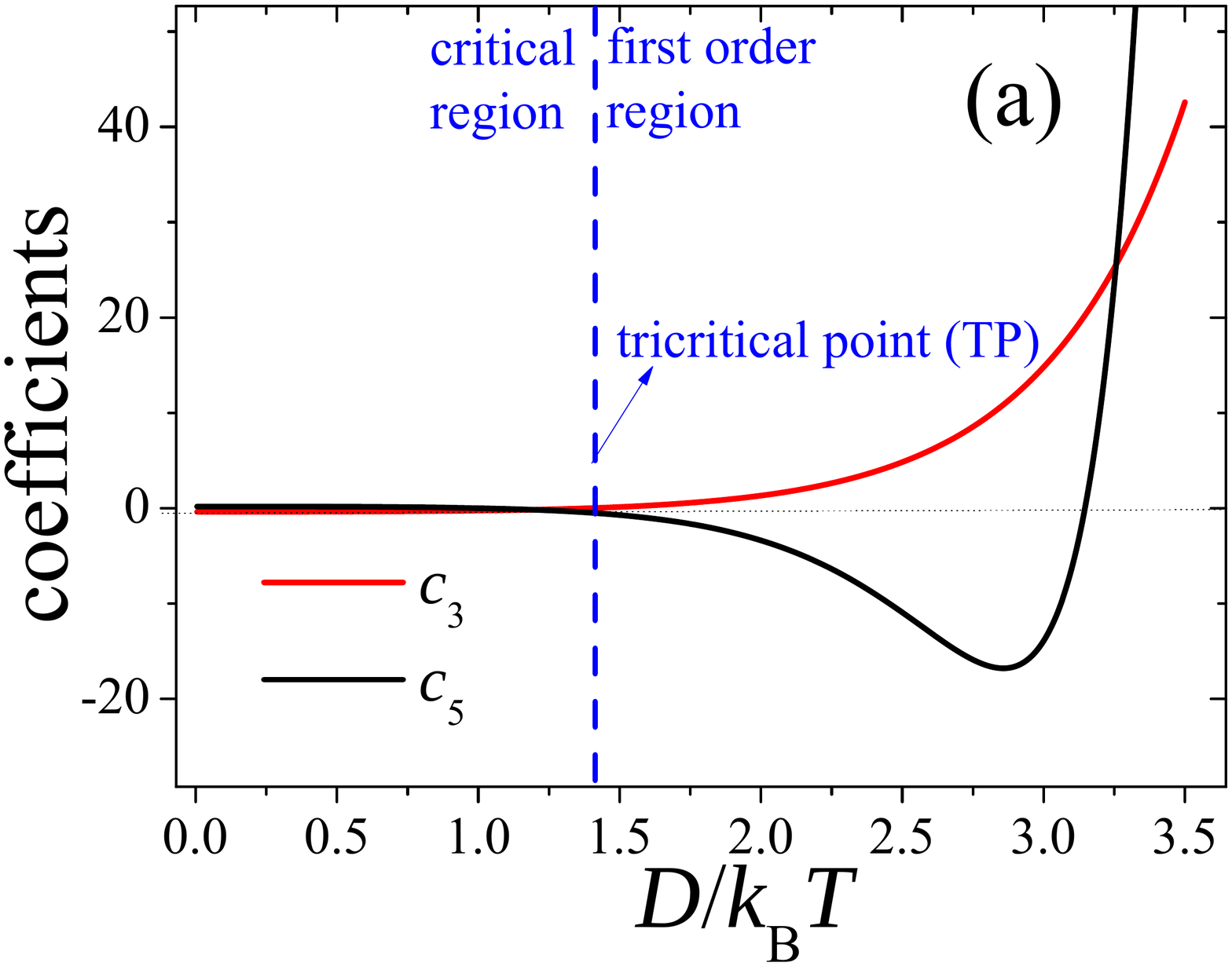} %
\includegraphics[width=1.0\columnwidth]{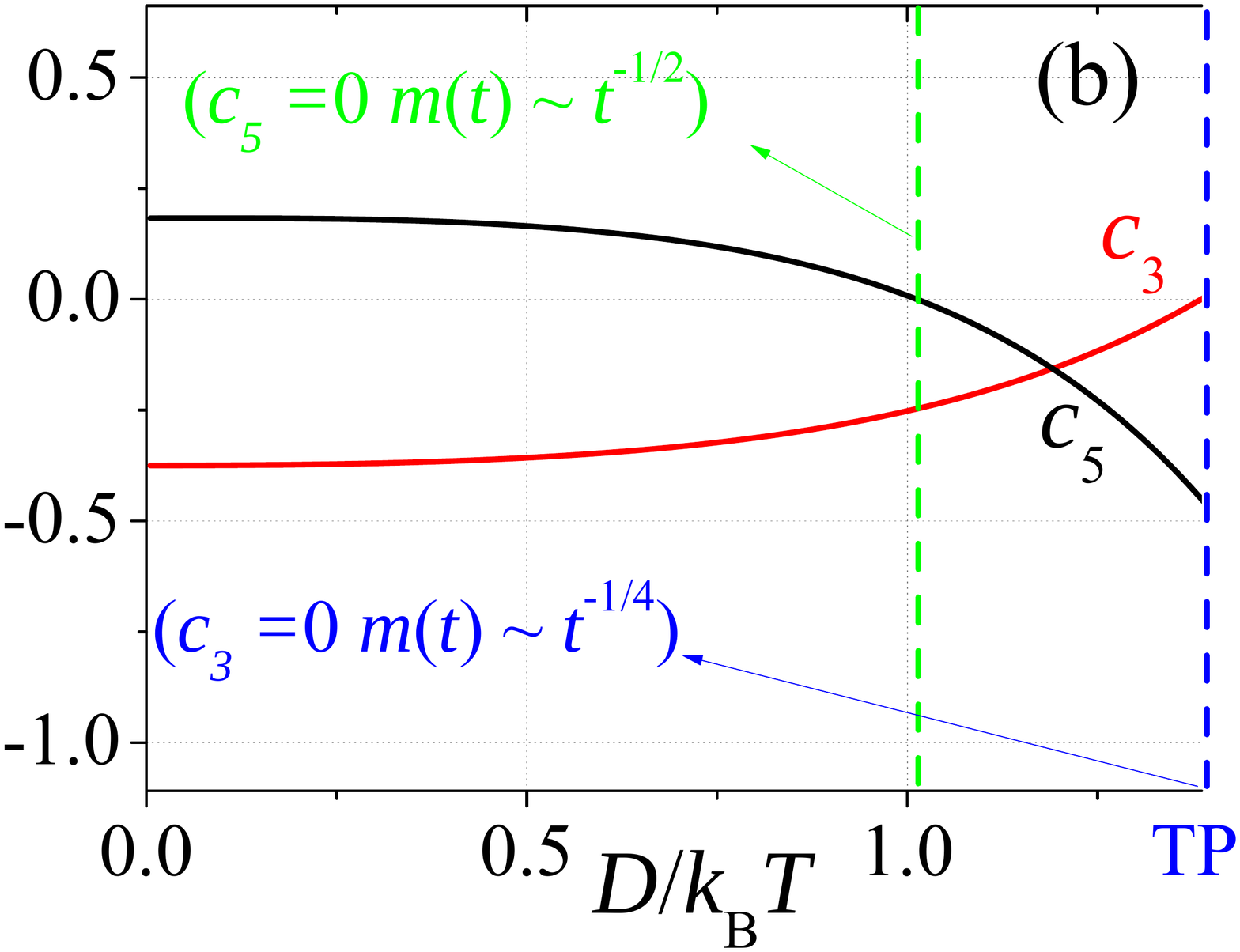}
\end{center}
\caption{(a) coefficients $c_{3}$ and $c_{5}$ of Eq: \protect\ref%
{Eq:differential_equation_MF_BC} as function of $D/k_{B}T$. (b) A zoom of
the critical region in (a). }
\label{Fig:coefficients}
\end{figure}

One can observe that values of $\lambda $ (second row in table \ref%
{Table:raw}) corroborate what we observed with the power-law predicted via
MF approximation. For the tricritical point:$\frac{D}{k_{B}T}=$ $2\ln 2$,
one obtains $\lambda =0.2665(1)$ in relation to theoretical value $\lambda =%
\frac{1}{4}$. For values of$\frac{D}{k_{B}T}$ from 0 until 0.84, the
exponents agree with Ising universality: $\lambda =\frac{1}{2}$, however,
this has no obligation to happen since $c_{3}$ and $c_{5}$ are not null in
this region as can be observed in Fig. \ref{Fig:coefficients}.

Fig. \ref{Fig:coefficients} (a) shows the behavior of these coefficients in
all range, while fig. \ref{Fig:coefficients} (b) shows a zoom of the
critical region. It is very interesting to observe that for $\frac{D}{k_{B}T}%
\leq 0.84$ when we observed that $\lambda \approx 1/2$, the coefficient
presents low variation when compared with the region after the exact value
when such exponent is expected: $\frac{D}{k_{B}T}=\ln (13-\sqrt{105})\approx
1.\,\allowbreak 0127$, dashed green line in Fig. \ref{Fig:coefficients} (b),
until TP (dashed blue line in this same figure). In addition we calculated
the exponent $\lambda $ for this extra value (not presented in table \ref%
{Table:raw}), which resulted in $\lambda =0.4963(4)$. \ Backing to the table %
\ref{Table:raw} one observes a smaller value of $\lambda $ for $\frac{D}{%
k_{B}T}=$ $1.12$ ($\lambda =0.4724(5)$). Such tendency must continue until
the minimal value of $\lambda $ that will happen in the last point of the
critical line: TP. For example in another extra point simulated (which also
is not presented in table \ref{Table:raw}), that one where $c_{3}=c_{5}$
(crossover between curves), resulting in $\frac{D}{k_{B}T}\approx
1.\,\allowbreak 191064$, we find $\lambda =0.4278(2)$, which corroborates
the crossover phenomenon from critical to tricritical point.

\begin{figure}[tbp]
\begin{center}
\includegraphics[width=1.0\columnwidth]{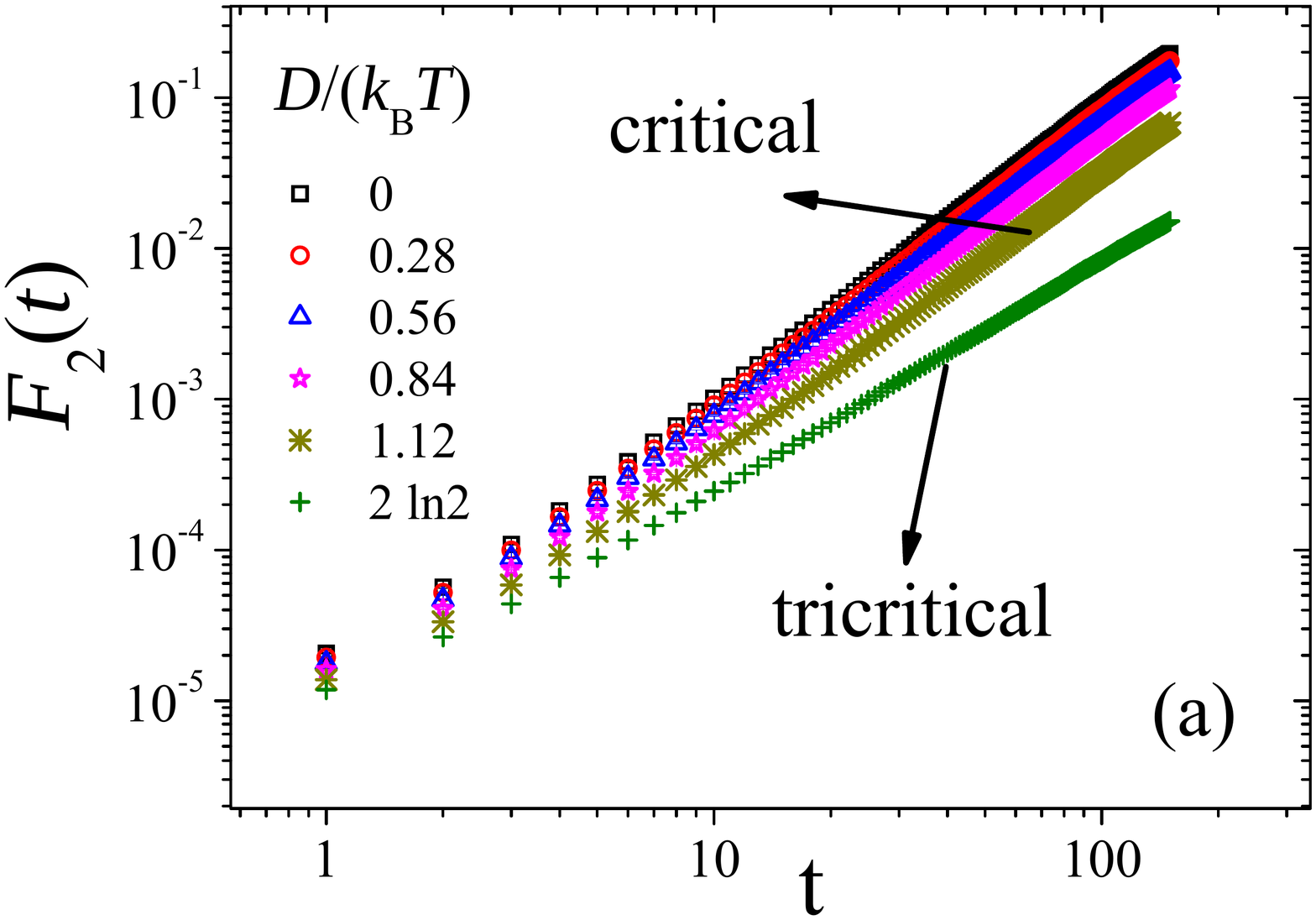} %
\includegraphics[width=1.0\columnwidth]{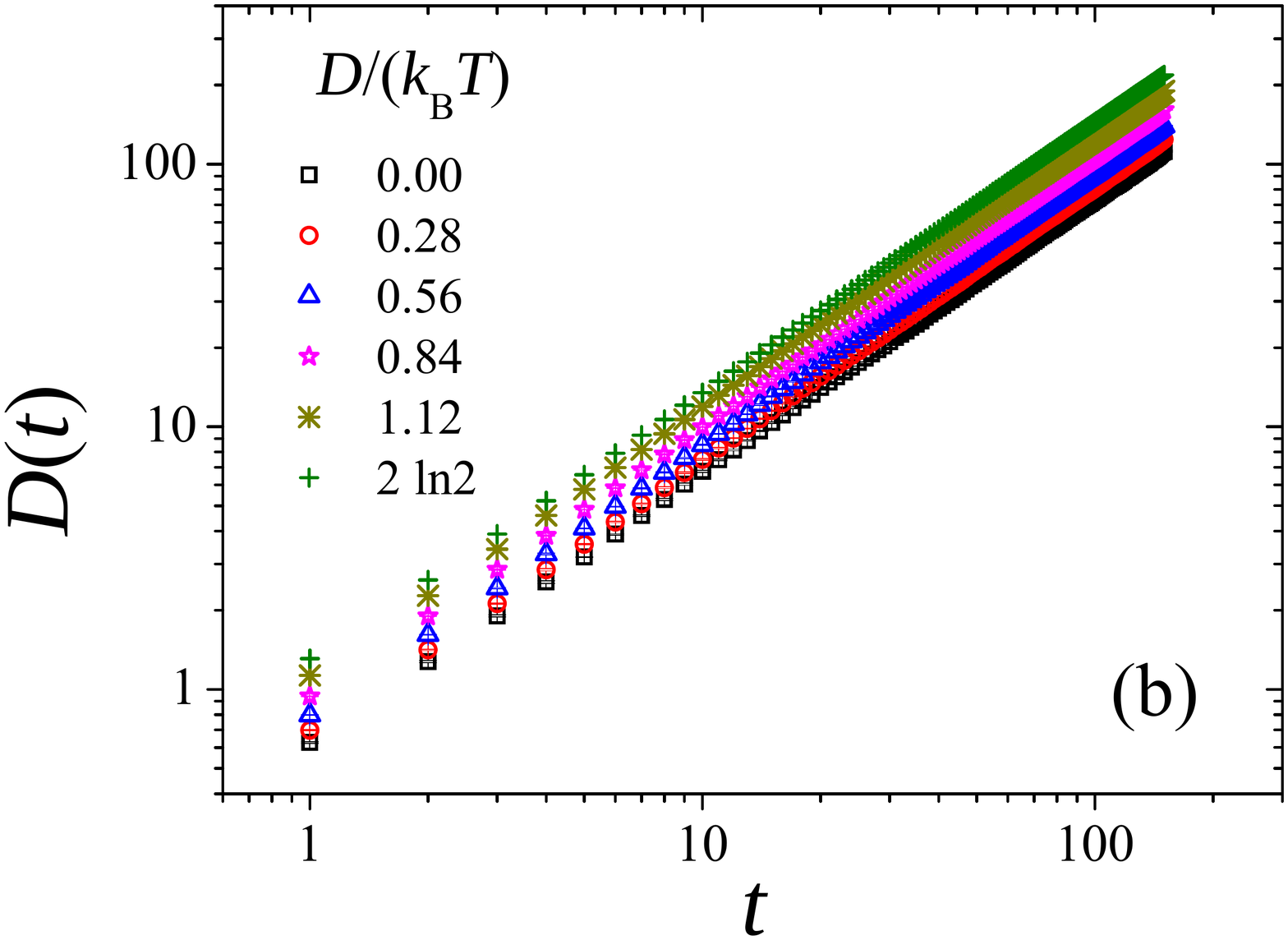}
\end{center}
\caption{(a) Time evolution of $F_{2}(t)$. (b) Time evolution of $D(t)$.
Both plots are presented for different values of $D/k_{B}T$:$\ 0$, $0.28$, $%
0.56$, $0.84$, $1.12$, and $2\ln 2$ (TP). }
\label{Fig:F2_and_D}
\end{figure}

In the following, we show in Figs. \ref{Fig:F2_and_D} (a) and (b) the time
evolution of $F_{2}(t)$ and $D(t)$, corresponding to exponents $\varsigma $
and $\eta $, for the same values of $\frac{D}{k_{B}T}$ used to plot the time
evolution of $m(t)$. For $F_{2}(t)$ (Fig. \ref{Fig:F2_and_D} (a) ) we can
observe a monotonic diminution of $\varsigma $ that leads to the minimal
value for the TP: $\varsigma =1.529(1)$, in addition the values of $%
\varsigma $ and $\vartheta $ are shown in table \ref{Table:raw}. On the
other hand, as suggested by the Fig. \ref{Fig:F2_and_D} (b) the exponent $%
\vartheta $ seems to remain approximately equal to 1 along the critical line
(see the values in table \ref{Table:raw}).

The critical exponents $\beta $, $z$, and $\nu $, obtained from raw
exponents, are shown in table \ref{Table:critical_exponents}.

\begin{table*}[tbp] \centering%
\begin{tabular}{lllllll}
\hline\hline
$\frac{D}{k_{B}T}$ & $0.00$ & $0.28$ & $0.56$ & $0.84$ & $1.12$ & $2\ln 2$
(TP) \\ \hline\hline
$\beta $ & $0.4982(5)$ & $0.4954(4)$ & $0.4948(4)$ & $0.4895(3)$ & $%
0.4658(5) $ & $0.2598(1)$ \\ 
$z$ & $2.050(2)$ & $2.047(3)$ & $2.051(3)$ & $2.049(2)$ & $2.109(2)$ & $%
1.962(1)$ \\ 
$\nu $ & $0.4946(6)$ & $0.4957(8)$ & $0.4979(8)$ & $0.5002(5)$ & $0.5198(5)$
& $0.4968(3)$ \\ \hline\hline
\end{tabular}%
\caption{Mean-field critical exponents from Blume Capel model along the critical line
obtained with TDMC simulations}\label{Table:critical_exponents}%
\end{table*}%

One can observe that classical Ising exponents were observed until $\frac{D}{%
k_{B}T}=0.84$. For $\frac{D}{k_{B}T}=1.12>1.\,\allowbreak 0127$ and thus,
above the point where $c_{5}=0$, the exponent presents a sensibility with
proximity to the tricritical point (crossover) and a decrease in $\beta $,
and an increase in $z$ and $\nu $ is observed. However, exactly in the TP
one observes a decrease in $z$, but yet $z\approx 2$ (spin 1/2 Ising) unlike
what happens in two dimensions (see \cite{SilvaPRE2002}), since one observes
an increase in $z$ when compared to the critical points. Here, it is
important to say that for TP, one used $d_{c}=3$, and for the critical
points one used $d_{c}=4$. The exponents $\beta $ and $\nu $ corroborates
the classical estimates for TP ($\beta =1/4$ and $\nu =1/2$) corroborating
that $d_{c}$ is indeed $3$ for TP.

Finally, we would like to revisit the discussion about the initial behavior
of the magnetization, in light of TDMC simulations in the MF regime. Both,
Ising-like and tricritical points in TDMC in two dimensions present a
power-law behavior $m(t)\sim t^{\theta }$ when the initial magnetization $%
m_{0}<<1$, but with different exponents $\theta >0$ for the first case and $%
\theta <0$ as we previously remembered at the beginning of this work. For
example, for Ising-like we found $\theta ^{(\text{Ising})}\approx 0.2$ while 
$\theta ^{(\text{TP})}$ $\approx -0.5$. An explanation of this fact could be
related to the global persistence phenomena in both situations. Global
persistence $P(t)$, is the probability of the magnetization remains positive
until time $t$ initially proposed by Majumdar et al. \cite{Majundar1996} and
that can find applications in systems from game theory until examples in
Econophysics \cite{SilvaPersextra}. Details in how to numerically calculate $%
P(t)$ in the context of MC simulations can be found for example \cite%
{SilvaPers}.

One knows that at the critical temperature, the persistence in
two-dimensional spin systems must behave as a power-law $P(t)\sim t^{-\theta
_{g}}$, where $\theta _{g}$ is the persistence exponent, which is valid for
simulations starting from a fixed (but random) $m_{0}<<1$. Thus, we
performed TDMC simulations in the MF regime to obtain $\theta _{g}$
comparing it with two-dimensional results.

\begin{figure}[tbp]
\begin{center}
\includegraphics[width=1.0\columnwidth]{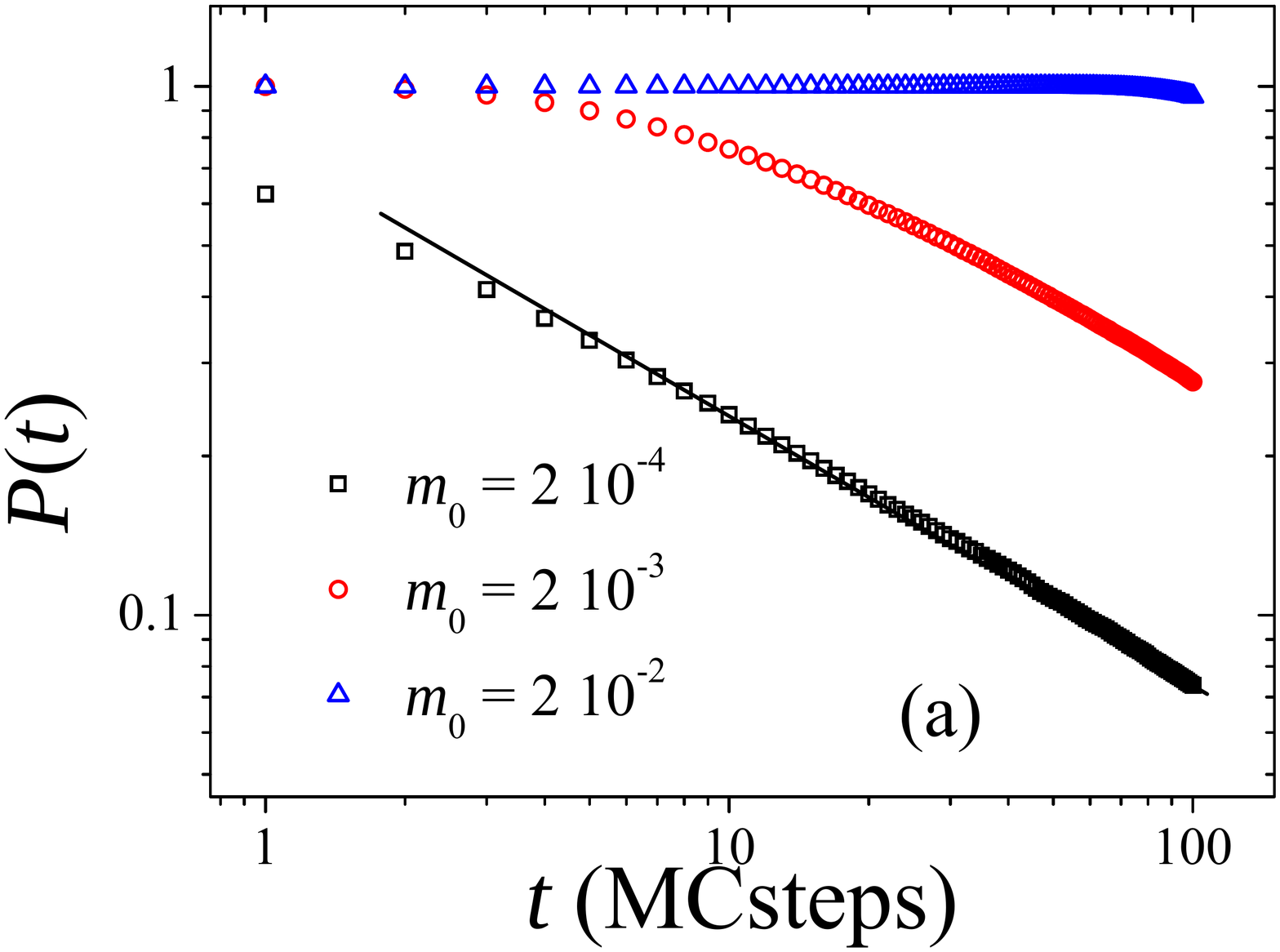} %
\includegraphics[width=1.0\columnwidth]{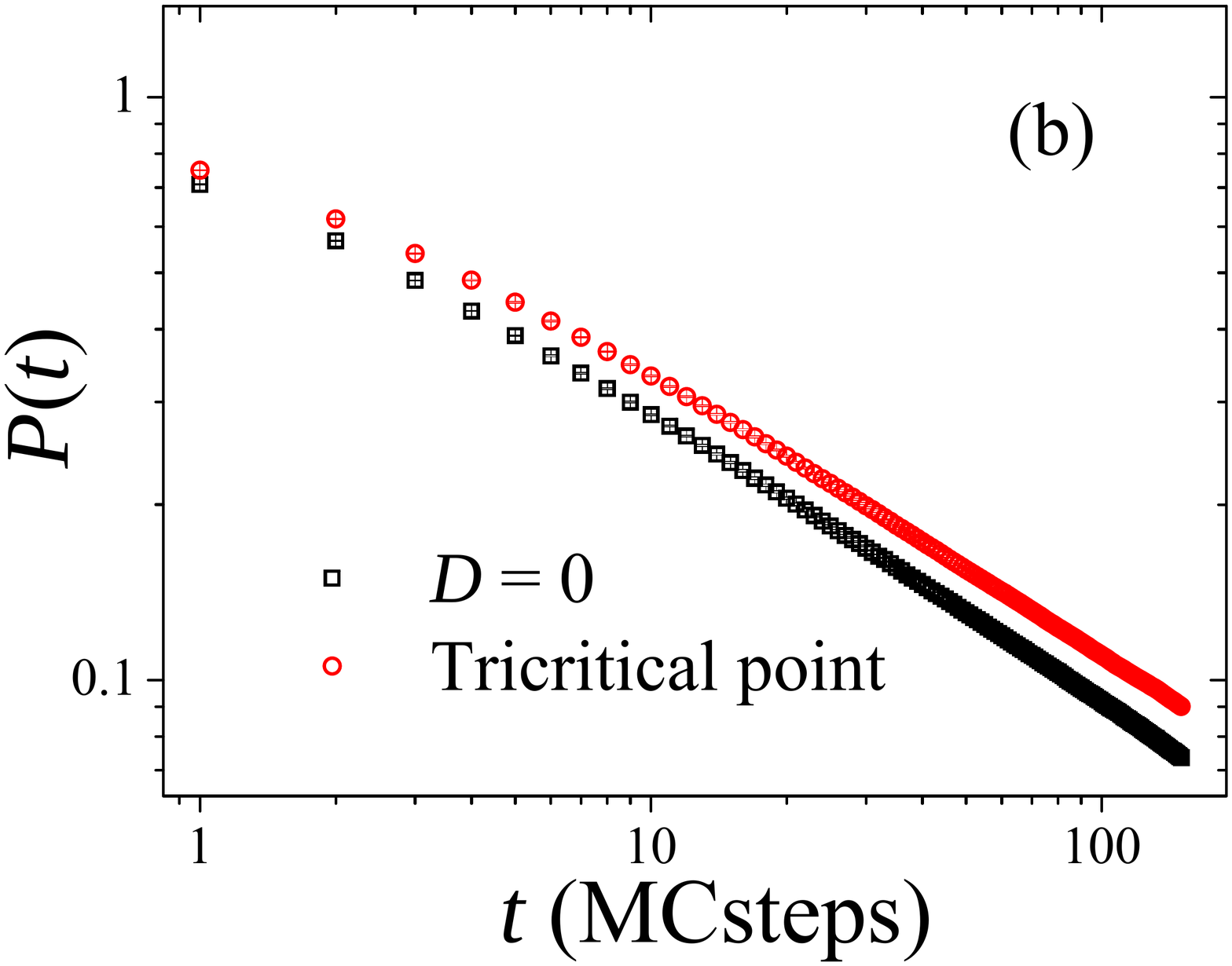}
\end{center}
\caption{Time evolution of persistence in the mean-field regime. (a) Ising
model for different values of $m_{0}$ (b) For a critical and for the
tricritical point of BC model when $m_{0}=10^{-4}$. }
\label{Fig:Persistence}
\end{figure}

Fig. \ref{Fig:Persistence} (a) shows the time evolution of $P(t)$ for Ising
model. We used different values of $m_{0}$, and a robust power law can be
observed for $m_{0}=10^{-4}$. We measure $\theta _{g}$ in this situation,
and one finds $\theta _{g}^{(\text{Ising-MF})}=0.542(2)$ which is very
different from one found for two-dimensional Ising model: $\theta _{g}^{(%
\text{Ising-2d})}\approx 0.238(3)$ \cite{Schulke}. This result corroborates
Majundar's result for the $n\rightarrow \infty $ limit of the O($n$) model: $%
\theta _{g}=1/2$ for $d>4$. Now, let us consider $\theta _{g}$ in the MF
regime for the BC\ model. Thus we consider two cases one critical, $\frac{D}{%
k_{B}T}=0$ and the tricritical $\frac{D}{k_{B}T}=2\ln 2$. The time
evolutions can be observed in Fig. \ref{Fig:Persistence} (b). We used in
both cases $m_{0}=10^{-4}$. For the critical point one obtained $\theta
_{g}^{(\text{BC-MF-Crit})}=0.5011(7)$, while a similar value is found for TP 
$\theta _{g}^{(\text{BC-MF-TP})}=0.488(3)$, differently from occurs for the
two dimensional BC $\theta _{g}^{(\text{BC-2d-Crit})}=0.241(4)$ and $\theta
_{g}^{(\text{BC-2d-TP})}=1.080(4)$. It is suggestive that $\theta _{g}^{(%
\text{BC-MF-Crit})}>\theta _{g}^{(\text{BC-2d-Crit})}$ and one observes an
initial increase of the magnetization characterized by a $\theta >0$ in the
case of critical points while that for the TP point: $\theta _{g}^{(\text{%
BC-MF-TP})}<\theta _{g}^{(\text{BC-2d-TP})}$ an initial decrease of the
magnetization characterized by a $\theta <0$ is observed. Short-range
systems at high temperature that are suddenly placed at $T=T_{c}$, presents
characteristics of a system in the MF regime that has a kind of inertia to
behave as a short-range system, therefore this attempt of the system into
establishing its new behavior leads to the initial anomalous behavior
observed from different ways for critical and tricritical points.

\section{Conclusions and summaries}

\label{Sec:Conclusions}

In this paper, we establish time-dependent Monte Carlo simulations in the
Mean-Field regime to study the relaxation of magnetization and other
quantities in Ising-like systems with spin 1/2 and 1. Differently from
short-range systems where appear an anomalous initial increase (critical) or
decrease (tricritical) of magnetization when properly prepared, such MF
systems are described by Eq. \ref{Eq:Initial_magnetization_behavior} for
spin 1/2 and by transcendental equation \ref{Eq:transcedental} for the
Blume-Capel model, this last one which can be analytically solved in some
particular cases. Our simulations corroborate such behavior and in addition,
we obtained the critical and tricritical parameters considering the
optimization of the expected power laws and we obtained the critical
mean-field exponents under the context of the crossover between the critical
and tricritical behavior. Finally, we explore the global persistence
phenomena for both: critical and tricritical points showing that mean-field
global persistence exponents, differently as occurs in two-dimensional
systems are similar for critical and tricritical points, which in our
opinion suggests a possible explanation to the differences between the
initial anomalous behavior of magnetization between critical and tricritical
points in two dimensions. \ 

\textbf{Acknowledgements} R. da Silva thanks CNPq for financial support
under grant numbers 311236/2018-9, 424052/2018-0, and 408163/2018-6.

\end{document}